\documentstyle[twoside,fleqn,espcrc2]{article}
\newcommand{\R}{\rm I\kern-.2emR}
\newcommand{\C}{\rm \kern.25em\vrule height1.4ex
 depth-.12ex width.06em\kern-.31em C}
\newcommand{\N}{{\rm I\kern-.16em N}}
\newcommand{\Z}{{\rm Z\kern-.35em Z}}
\newcommand{\bee}{\begin{equation}}
\newcommand{\ee}{\end{equation}}
\newcommand{\ba}{\begin{array}}
\newcommand{\ea}{\end{array}}
\newcommand{\bea}{\begin{eqnarray}}
\newcommand{\eea}{\end{eqnarray}}

\newcommand{\AmS}{{\protect\the\textfont2
  A\kern-.1667em\lower.5ex\hbox{M}\kern-.125emS}}

\title{The Perturbative Method Fails in Non-Abelian Models
\thanks {Talk given by E.~Seiler}}
\author{A. Patrascioiu\address{Physics Department
         \\ University of Arizona, Tucson, AZ 85721}%
\ and E.~Seiler\address{Max-Planck Institut f\"ur Physik\\
 -Werner-Heisenberg Institut-\\
 F\"ohringer Ring 6, D-80805 M\"unchen,
 F.R.Germany}
}

\begin{document}

\begin{abstract}
It is shown that perturbation theory in
$2D$ nonlinear $\sigma$-models as well gauge theories in dimension
$D\geq 2$ produces answers that depend on boundary conditions even after
the infinite volume limit has been taken. This unphysical phenomenon occurs
only in the non-Abelian versions of those models, starting at $O(1/\beta^2)$.
It is not present in the true (nonperturbatively
defined) models and represents a failure of the perturbative method.
It is related to a hitherto unnoticed type of low-lying excitation,
dubbed super-instanton, that dominates the low-temperature (= weak
coupling) regime of these models.
\end{abstract}

\maketitle

\vskip4mm \noindent
\section{Introduction}
\vskip2mm
At the conference {\it Lattice 92} in Amsterdam we argued \cite{Lat92}
on the basis of percolation theoretic considerations that all $2D$
$O(N)$ models must have at low temperature a soft spin wave phase
characterized by power-like decay of correlations. This argument, while
not fully rigorous, is intuitively compelling and was not challenged by
anybody. Maybe the most striking conclusion to be drawn from
it is that asymptotic scaling has to fail and that the
perturbatively computed $\beta$-function does not describe
correctly the variation of the correlation length with the
bare coupling or its inverse $\beta$.

How can that be? Only if, as we have been arguing for many years
\cite{Pat,FF}, perturbation theory (PT) does not produce the correct
asymptotic expansion for expectation values in the infinite volume limit
in those models.
Several colleagues questioned this possibility, arguing that such a
failure of PT should show up in PT itself. While this argument
is not logically compelling, it induced us nevertheless to search for
possible pathologies in PT. Here I want to report that we indeed found
such a pathology, and in fact not only in $2D$ non-Abelian nonlinear
$\sigma$-models but also in non-Abelian lattice Yang-Mills (LYM) models:

{\it PT produces results that remain sensitive to boundary
conditions (b.c.) even in the infinite volume limit, whereas the true
(nonperturbatively defined) models do not show such a dependence.}

\section{The Low-Lying Excitations}

In order to motivate the b.c. we are studying, we first have to
look at the low-lying excitations of these models, which will
be relevant for the behavior of the Gibbs state at low temperature
(= weak coupling).

Let us first look at the $2D$ $O(N)$ models. Since the work of
Kosterlitz and Thouless \cite{KT} it has been assumed that for
$N=2$ the crucial excitations are the vortices, which in isolation
have an energy $O(\ln L)$ ($L$ is the linear size of the system), and
which therefore will be bound into pairs at low temperature.
Kosterlitz and Thouless also argued that for $N=3$ the relevant
excitations are the instantons (even though that name was coined
later), which have an energy $O(L^o)$. But there is a different type
of excitation which has arbitrarily low energy and which we therefore
dubbed `super-instanton'. It will truly dominate at low temperature.

The super-instantons can be loosely described as follows:
Fix a spin at the origin to a direction $e_o\in S^{N-1}$ and fix the
spins at the boundary to a (generally different) direction $e_1$,
forming an angle $\phi_o$ with $e_o$.
The super-instanton is then the configuration of lowest energy
satisfying these b.c.. It describes spins $s_i$ that turn gradually
from $e_o$ to $e_1$ in such a way that the angle between $e_o$ and
$s_i$ satisfies the lattice Laplace equation (as does the KT vortex).
It is not hard to convince oneself that the energy $E$ of such a
super-instanton will be $O(\phi_o^2/\ln L)$.

Analogous configurations can also be constructed in other dimensions.
In $1D$ the energy will be $O(1/L)$, whereas for $D\geq 3$ the
minimal energy will satisfy $E\geq E_o>0$.

The fact that in $D\leq 2$  the energy can be made arbitrarily small
is responsible for the Mermin-Wagner theorem \cite{MW}, which
forbids spontaneous symmetry breaking (SSB): the abundance of
super-instantons will disorder the system. In $D\geq 3$, on the
other hand, super-instantons will be strongly suppressed at
low temperature, and SSB will occur.

How about LYM models? It is well known that in the
axial gauge ($U_{i,i+e_\mu}=1$ for $\mu=1$), LYM theory looks like a
collection of $1D$ spin chains,
where for $D\geq 3$ the chains are coupled via the plaquettes
orthogonal to the 1-direction. The super-instantons in this
case are simply the super-instantons of the corresponding spin
chains. Of course they can also be described in a gauge invariant
way as configurations which have a $L\times 1$ Wilson loop
fixed to a (generally nontrivial) value. These super-instantons
have an energy $O(1/L)$ in all $D\geq 2$ and enforce the analogue
of the Mermin-Wagner theorem in gauge theories, which can be stated
as the absence of SSB in the complete axial gauge and has been proven
long ago \cite{SY,Luscher,Pat}.

\begin{table*}[hbt]
\noindent
{\bf Tab.1:} {\it The second order PT coefficients of the energy
for $2D$ $O(N)$ models with super-instanton b.c.. For $O(2)$ we give
$32c_2(L)$, for $O(3)$ $16c_2(L)$. The last column (labeled `p.b.c.')
gives the infinite volume limit obtained using periodic b.c., \cite{Has}.}

  \vbox{\offinterlineskip\halign{
  \strut#&&\quad $#$\hskip3pt&#\cr
   &L&& 20&& 30&& 40&& 60&& 80&& 120&& 160&& 240&&\infty&&  && p.b.c.&\cr
   \noalign{\hrule\vskip1pt\hrule}
   &O(2)&&.7973&&.8161&&.8265&&.8384&&.8455&&.8544&&.8599&&.8670
   &&1.0&& &&1.0&\cr
   \noalign{\hrule}
   &O(3)&&1.1283&&1.1852&&1.2171&&1.2539&&1.2759&&1.3027&&1.3193&&1.3401
   &&1.6663&& &&1.0\cr
   }}

\medskip
{\bf Tab.2:} {\it The PT coefficients $c_2(L)$ for the energy of the
plaquette $P_1$ computed with super-instanton b.c. in LYM models with
gauge group $U(1)$ and $SU(2)$. In the last column we give the infinite
volume limits of the energy obtained with periodic b.c. \cite{WWW}.}

  \medskip
  \vbox{\offinterlineskip\halign{
  \strut#&&\quad $#$\hskip5pt&#\cr
   &L&& 8&& 10&& 12&& 16&& 20&& 30&&\infty&& &&p.b.c.&\cr
   \noalign{\hrule\vskip1pt\hrule}
   &U(1)&&.03834&&.04152&&.04369&&.04648&&.04820&&.05056
   &&1/18&& &&1/18&\cr
   \noalign{\hrule}
   &SU(2)&&.2536&&.2852&&.3061&&.3320&&.3472&&.3669&&.4063
   &&  &&.2325&\cr
   }}

\end{table*}

The central point is that both in $2D$ $O(N)$ and in LYM
models there are large fluctuations present at all values of the
bare coupling and that these fluctuations make PT untrustworthy.

\section{Perturbation Theory}

In a fixed volume $L^D$, for large $\beta$, PT produces an
asymptotic expansion in $1/\beta$ for quantities such as the energy:

\bee
E(L)=1-{c_1(L)\over\beta}-{c_2(L)\over\beta^2}+O({1\over\beta^3})
\ee
The conventional procedure is to take the limits
$\lim_{L\to\infty}c_j(L)$ and hope that they will be the
coefficients of the (unique if it exists) asymptotic expansion in the
infinite volume limit. The point made here is that this
procedure is ambiguous, because the limits depend on the b.c. used.
For $1D$ $O(N)$ models this is a well known fact \cite{BR}, but the belief
has been \cite{Has} that for $2D$ spin models or LYM theory such an effect
does not occur.

We will see this effect when we compare the conventional
answers, obtained with periodic b.c., with those obtained with
the so-called super-instanton b.c. (s.i.b.c.); the latter are motivated
by the fact that it should be as legitimate to do PT around a
super-instanton (which is a local minimum of the energy) as to do it
around the trivial, completely ordered ground state.

For computational reasons we only use trivial s.i.b.c., where the
turning angle $\phi_o$ in the spin models is $0$ and the long thin
Wilson loop in LYM theory is fixed to the identity.
It is important to note that these are legitimate b.c. even though we fix a
variable in the center of the lattice, because of the
Mermin-Wagner theorem \cite{MW}
(or its analogue in LYM \cite{SY,Luscher,Pat}): Fixing one spin
(or fixing one link variable) does not change any expectation values
of invariant quantities. Fixing then in addition the boundary variables
is clearly a b.c. and will leave no effect in the
thermodynamic limit.

We computed with s.i.b.c. the PT coefficients $c_2(L)$ of order
$1/\beta^2$ for the energy $E(L)$ in the $2D$ $O(N)$ models and in the
$3D$ LYM model with gauge groups $U(1)$ and $SU(2)$,
and compared them with the results obtained using periodic
boundary conditions (p.b.c.).
In the $O(N)$ model $E=\langle s_o\cdot s_1\rangle$, whereas in the
LYM model $E=1/N \langle\rm{tr}U_{P_1}\rangle$,
where the plaquette $P_1$ is in the
center of lattice, at the end of the Wilson loop that was fixed by the
s.i.b.c.. We find the following:

\noindent
(1) Abelian models:
\bee\lim_{L\to\infty}c_2^{s.i.b.c.}(L)=\lim_{L\to\infty}c_2^{p.b.c.}(L)
\ee
(2) Non-Abelian models:
\bee\lim_{L\to\infty}c_2^{s.i.b.c.}(L)\neq\lim_{L\to\infty}c_2^{p.b.c.}(L)
\ee
In other words, PT fails for non-Abelian models, at least with
some b.c.!

The computation was done, following a suggestion by A.Sokal \cite{Sokal},
with the use of an eigenfunction expansion. The numbers are given in
Tab.1 for
the $2D$ $O(2)$ and $O(3)$ models (for general $O(N)$ models see
\cite{si}) and in Tab.2. for $3D$ LYM with gauge groups $U(1)$ and $SU(2)$.

The $L\to\infty$ limits for the $2D$ models are obtained by fitting the
data to a 3rd order polynomial in $1/\ln L$, which represents the numbers
to all the digits given. For the LYM models the infinite volume
limits are obtained by fitting to a 3rd order polynomial in $1/L$
which again represents the data to all digits given.
For the spin models Niedermayer and Weisz \cite{NW} confirmed
Fact (2) above, going to even larger lattices.

A final remark concerns the $\beta$-function. Of course if we find such
an effect for a short distance quantity such as the energy, it should
not come as a surprise that also long range quantities are affected,
in particular the $\beta$-function. We verified this for the $2D$
$O(N)$ model.

Our findings are described in more detail in two papers \cite{si} and
\cite{sig}.



\begin{thebibliography}{99}
%
\bibitem{Lat92}A.Patrascioiu~and~E.Seiler,~{\sl Nucl.Phys.B. (Proc.
Suppl.)} {\bf 30} (1993) 184.
\bibitem{Pat}A.Patrascioiu, {\sl Phys.Rev.Lett.} {\bf 54} (1985) 2292.
\bibitem{FF}A.Patrascioiu and E.Seiler, {\it The Difference between Abelian
and Non-Abelian Models: Fact and Fancy}, preprint MPI-Ph/91-88.
\bibitem{KT}J.M.Kosterlitz and D.J.Thouless, {\sl J. Phys. (Paris)}
{\bf 32} (1975) 581.
\bibitem{MW}N.D.Mermin and H.Wagner, {\sl Phys. Rev. Lett.} {\bf 17} (1966)
 1133;
N.D.Mermin, {\sl J. Math. Phys.} {\bf 8} (1967) 1061.
\bibitem{SY}B.Simon and L.G.Yaffe, {\sl Phys.Lett.}
{\bf 115B} (1982) 145.
\bibitem{Luscher}M. L\"uscher, {\it Absence of Spontaneous Gauge
Symmetry Breaking in Hamiltonian Lattice Gauge Theories}, preprint DESY-77/16.
\bibitem{BR} Y.Brihaye and P.Rossi, {\sl Nucl.Phys.}
{\bf B235} (1984) 226
%
\bibitem{Sokal}A.Sokal, private communication.
\bibitem{NW}F.Niedermayer and P.Weisz, private communication.
\bibitem{si}A.Patrascioiu and E.Seiler, {\it Super-Instantons and
the Reliability of Perturbation Theory in Non-Abelian Models},
$\langle$hep-lat 9311019$\rangle$, to appear in Phys.Rev.Lett.
\bibitem{sig}A.Patrascioiu and E.Seiler, {\it Super-Instantons in Gauge
Theories and Troubles with Perturbation Theory},
$\langle$hep-lat 9402003$\rangle$, to appear in Phys.Rev.Lett.
\bibitem{Has}P.Hasenfratz, {\sl Phys.Lett.}{\bf B141} (1984) 385.
\bibitem{WWW}R.Wohlert, P.Weisz and W.Wetzel
{\sl Nucl.Phys.} {\bf B 259} (1985) 85.
%
\end{thebibliography}
\end{document}